# Surface-Enhanced Raman Spectroscopy of Graphene Integrated in Plasmonic Silicon Platforms with Three-Dimensional Nanotopography

*Maria Kanidi*[1,2], *Alva Dagkli*[3], *Nikolaos Kelaidis*[4,5], *Dimitrios Palles*[1], *Sigiava Aminalragia-Giamini*[4,6], *Jose Marquez-Velasco*[4,7], *Alan Colli*[8], *Athanasios Dimoulas*[4], *Elefterios Lidorikis*[3], *Maria Kandyla*[1,*], *Efstratios I. Kamitsos*[1]

[1]Theoretical and Physical Chemistry Institute, National Hellenic Research Foundation, 48 Vassileos Constantinou Avenue, 11635 Athens, Greece

[2]Department of Material Science, University of Patras, University Campus, 26504 Rio, Greece

[3]Department of Materials Science and Engineering, University of Ioannina, 45110 Ioannina, Greece

[4]Institute of Nanoscience and Nanotechnology, National Center for Scientific Research 'Demokritos', 15310 Athens, Greece

[5]Faculty of Engineering, Environment and Computing, Coventry University, Priory Street, Coventry CV1 5FB, UK

[6]Department of Physics, University of Athens, Zografou University Campus, 15784 Athens, Greece

[7]Department of Physics, National Technical University of Athens, 9 Iroon Polytechniou st., 15780 Athens, Greece

[8]Emberion Ltd, Unit 151 Cambridge Science Park, Cambridge CB4 0GN, UK

**ABSTRACT**

Integrating graphene with plasmonic nanostructures results in multifunctional hybrid systems with enhanced performance for numerous applications. In this work, we take advantage of the remarkable mechanical properties of graphene to combine it with scalable 3D plasmonic nanostructured silicon substrates, which enhance the interaction of graphene with electromagnetic radiation. Large areas of femtosecond laser-structured arrays of silicon nanopillars, decorated with gold nanoparticles, are integrated with graphene, which conforms to the substrate nanotopography. We obtain Raman spectra at 488, 514, 633, and 785 nm excitation wavelengths, spanning the entire visible range. For all excitation wavelengths, the Raman signal of graphene is enhanced by 2–3 orders of magnitude, similarly to the highest enhancements measured to date, concerning surface-enhanced Raman spectroscopy (SERS) of graphene on plasmonic substrates. Moreover, in contrast to traditional deposition and lithographic methods, the fabrication method employed here relies on single-step, maskless, cost-effective, rapid laser processing of silicon in water, amenable to large-scale fabrication. Finite-difference time-domain simulations elucidate the advantages of the 3D topography of the substrate. Conformation of graphene to the Au-decorated silicon nanopillars enables graphene to sample near fields from an increased number of nanoparticles. Due to synergistic effects with the nanopillars, different nanoparticles become more active for different wavelengths and locations on the pillars, providing broadband enhancement. Nanostructured plasmonic silicon is a promising platform for integration with graphene and other 2D materials, for next-generation applications of large-area hybrid nanomaterials in the fields of sensing, photonics, optoelectronics, and medical diagnostics.

**Introduction**

Graphene is the celebrated two-dimensional material with remarkable mechanical, electronic, thermal, chemical, and optical properties[1,2] and a wide range of applications in the fields of sensing, electronics and optoelectronics, light harvesting, and photonics, among others[2-5]. Most of these applications involve the interaction of graphene with electromagnetic fields, therefore enhancement of this interaction is essential. Integrating the atomically thin graphene with plasmonic nanostructures results in graphene experiencing enhanced electromagnetic near fields due to coupling with surface plasmon modes[6]. Due to its physical and chemical properties, graphene is particularly attractive for integration, resulting in multifunctional hybrid systems with enhanced efficiency for numerous applications[7]. Specifically, plasmonic nanostructures in combination with graphene or other two-dimensional materials have already demonstrated a promising potential for the development of high-performance photodetectors[5,8], solar cells[7], optical modulators[9], fuel cells[10], as well as chemical and biological sensors[11,12].

A revealing testbed for the understanding and exploitation of the interaction of graphene with plasmonic substrates is surface-enhanced Raman scattering (SERS), which relies fundamentally on light-graphene interaction. Enhancing the Raman spectrum of graphene is important for understanding the behavior of graphene on a given substrate, as it allows for the identification of the number of layers, type of doping, strain, defects, temperature effects, chemical modification, disorder, and edges, among others[13-15]. The vital dependence of the properties of 2D materials on the underlying substrate and the modulation of these properties by substrate engineering have been reviewed recently[16]. Integration of graphene with plasmonic metallic nanostructures has been shown to enhance the Raman signal of graphene by 2–3 orders of magnitude[17-28]. To this end, various kinds and geometries of metallic nanostructures have been employed, including nanodisks[23], nanodots[24], nanopyramids[22], polygons, dendrites, dense clusters[27], and irregular islands[18,27], which resulted in measured enhancement factors of the Raman signal of graphene as high as 1000. However, most of these structures, usually developed with multi-step and elaborate deposition and lithographic methods, are limited to laboratory-size areas which lack scalability, constricting real-world applications[29-31]. Furthermore, they are mostly restricted to 2D topographies, usually lying on flat substrates, with which graphene behaves mainly as a rigid over- or underlayer. Therefore, the challenge remains for integration of graphene with large-scale plasmonic nanostructures, taking advantage of the remarkable mechanical stability of graphene in 3D configurations.

Surface nanopatterning is an alternative to lithographic methods for the development of plasmonic nanostructures. Laser processing is a single-step, maskless, tabletop method to create uniformly micro/nanopatterned surfaces over large areas. Coating these micro/nanopatterned surfaces with thin metallic layers results in the spontaneous formation of metallic nanoparticles, rendering them plasmon-active. Nanosecond and femtosecond laser-patterned silicon wafers, coated with metallic thin films and nanoparticles, have been used as efficient SERS substrates with liquid analytes recently[29-36]. Nanostructured silicon with gold nanoparticles was employed as a SERS substrate for Rhodamine 6G spectroscopy and provided enhancement factors of $10^4$[31] and $10^7$[33,34]. Arrays of silicon nanopillars with silver nanoparticles demonstrated high sensitivity and reusability as a SERS substrate for Rhodamine 6G and Methylene blue[35], while resulting in a Raman enhancement factor of $10^9$ for melamine molecules[36] and $10^7$ for benzenethiol[30]. Arrays of silicon microsquares containing aggregates of silicon nanoparticles were coated with a thin silver film and reached a Raman enhancement factor of $10^6$ for 4-methylbenzenethiol molecules[29]. These substrates demonstrate not only large enhancement factors but also high uniformity of the Raman signal and enhancement factor over macroscopic areas[29,30,35]. They are robust with great structural stability, for long-term integration with spectroscopic devices having little degradation in performance[30]. More importantly, their 3D topography provides an increased surface area for nanoparticle deposition and pronounced plasmonic effects[29,33,34]. Indeed, gold nanoparticles on nanostructured silicon substrates show higher enhancement factors for Raman spectroscopy compared with gold nanoparticles on pristine silicon[31].

In this work, we take advantage of the remarkable mechanical properties of graphene to combine it for the first time with scalable 3D plasmonic silicon-based platforms, which enhance the interaction of graphene with electromagnetic radiation. As a non-rigid 2D solid material, graphene is ideal for integration with a 3D substrate due to its extraordinary strength, which allows it to conform to the substrate nanotopography and maintain its structural integrity. In more general terms, we exploit the unique ability of 2D materials to be combined with 3D substrates, as they can be grown over large areas with low-cost methods such as chemical vapor deposition, unlike bulk semiconductors which require high-quality epitaxial growth. Furthermore, the surface of graphene is chemically passivated, supporting integration with an arbitrary substrate[5]. We demonstrate the plasmonic enhancement of the interaction of graphene with the electromagnetic field via the SERS case, by employing large areas of femtosecond laser-structured arrays of silicon nanopillars, decorated with gold nanoparticles, as SERS substrates. We probe graphene for its plasmonic-enhanced Raman spectral signal at four excitation wavelengths, which span the visible range of the electromagnetic spectrum (488, 514, 633, and 785 nm). The Raman signal of graphene on the plasmonic 3D substrate is enhanced by 2–3 orders of magnitude, compared with pristine silicon substrates, which allows for probing the fine structure of Raman scattering in graphene. This enhancement is similar to the highest ones measured to date, concerning SERS of graphene on plasmonic substrates, and it is achieved via single-step, cost-effective, rapid laser nanopatterning methods, amenable to large-scale fabrication. Broadband enhancement is observed for all excitation wavelengths, across the entire visible electromagnetic spectrum. Numerical simulations employing the finite-difference time-domain (FDTD) method confirm the plasmonic nature of the Raman enhancement and elucidate the advantages of the 3D topography of the substrate. The Raman signal is calculated to increase monotonically with the degree of graphene conformation to the substrate, demonstrating the advantages of exploiting a 3D topography. The broadband and intense enhancement of the interaction of graphene with electromagnetic radiation, shown here, is pertinent to ultrasensitive sensors as well as to a broader family of silicon-integrated photonic and optoelectronic applications based on 2D materials, such as photodetectors, solar cells, light emitting diodes, optical modulators, *etc*.

**Methods**

Substrate preparation. Nanostructured silicon substrates were fabricated by femtosecond laser processing in water. The output pulse train of a regenerative Ti:sapphire laser amplifier (800 nm wavelength, 160 fs pulse duration, 20 KHz repetition rate) was frequency-doubled to a wavelength of 400 nm using a BBO crystal. The pulse train was focused on a p-type (100) silicon wafer to an average fluence of about 1 J/cm$^2$ on its surface, with the silicon wafer being placed in a cuvette filled with distilled water. The cuvette was mounted on a computer-controlled set of xy translation stages and raster scanned at an appropriate speed so that each spot on the silicon surface was irradiated by 1000 pulses. This process resulted in the formation of a quasi-ordered array of nanopillars on the silicon surface. The nanostructured silicon substrates were coated by thermal evaporation with a nominal 50-nm thick gold layer, resulting in a dispersion of gold nanoparticles on the surface[37,38]. In order to separate effects arising from the substrate geometry from plasmonic effects, three other types of substrates were also employed: a) nanostructured silicon without gold nanoparticles, b) flat, unstructured silicon with a 50-nm thick gold layer, and c) flat, pristine silicon without a gold layer.

Graphene growth and transfer (Fig. 1). Monolayer graphene was grown on a catalytic copper foil substrate by chemical vapor deposition (CVD), performed in an Annealsys commercial cold-wall reactor system. Before growth, the copper substrate was annealed in an argon and hydrogen atmosphere at 975$^o$C, to reduce the surface and remove the surface oxide. Following this step, graphene was grown at the same temperature using methane as the carbon precursor, diluted in argon. As-grown graphene was characterized by Raman spectroscopy and measurements were found to be typical of a single layer. The graphene film was transferred on the substrates used in this work (flat pristine silicon, flat silicon with 50-nm gold film, uncoated nanostructured silicon, and nanostructured silicon with gold nanoparticles), employing a standard graphene transferprocess[39]. First, the graphene/Cu sample was spin-coated with polymethylmethacrylate (PMMA). Copper was then etched in a FeCl3 solution and the membrane of PMMA-graphene was rinsed and ready to be transferred on the target substrate. In order to transfer graphene successfully, the target surface must be hydrophilic. Because nanostructured silicon with gold nanoparticles is hydrophobic, it was immersed briefly (~20 sec) in a mild piranha solution before graphene transfer, in order to render it hydrophilic without removing or damaging the gold nanoparticles. The uncoated flat and nanostructured silicon substrates were also cleaned in a piranha solution before graphene transfer. Finally, the PMMA-graphene membrane was transferred on the target substrate, which was then immersed in acetone to remove PMMA.

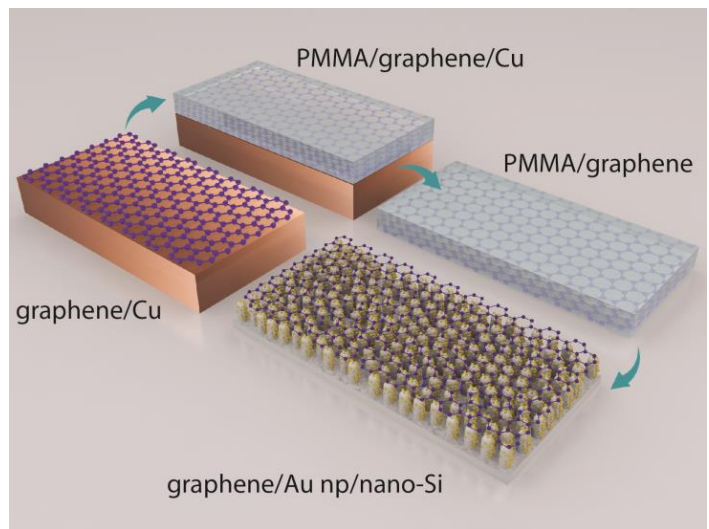

**Figure 1**: Schematic of graphene growth on copper and transfer on nanostructured silicon substrate with gold nanoparticles.

Micro-Raman spectroscopy setup. Raman spectra were acquired with a Renishaw inVia Reflex Raman microscope, equipped with a Peltier-cooled charge coupled device (CCD) and a motorized xyz microscope stage with a lens of magnification ×50, in a backscattering geometry. The 488 nm and 514.5 nm lines of an argon laser, the 633 nm line of a He-Ne laser, and the 785 nm line of a solid state diode laser were used for excitation. Together with the rest of the system configuration (grating, slit width, CCD partition) this results in a spectral resolution of ~1 cm$^{-1}$. The laser beam was focused on the sample to a spot of diameter 2–4 μm, depending on the laser wavelength and the scattering conditions. We employed various excitation laser powers, always keeping the incident power low enough to avoid sample damage and heating. Specifically, for the 488 nm, 514.5 nm, 633 nm, and 785 nm lines the power level employed on the sample was 1.72 mW, 0.39 mW, 1.32 mW, and 0.14 mW, respectively. For the nanostructured silicon substrates, care was taken to use the same substrate area for Raman spectra acquisition for all laser lines, to minimize spectral variations within the substrate. Raman measurements on the flat silicon substrate covered with a 50-nm gold film, with or without graphene, using the 785 nm excitation wavelength, show a broad asymmetric band centered at ca. 1450 cm$^{-1}$ (or ca. 885 nm) with a linewidth of ca. 500 cm$^{-1}$ (Supporting Information Fig. S1). Since this band is absent when using the other three excitation wavelengths, it should be a fluorescence rather than a vibrational band. Similar fluorescence bands have been observed in the Raman spectra of soda-lime silicate glasses with 785 nm excitation and they have been attributed to the presence of rare earth ions, like neodymium, in ppm concentrations in the soda-lime silicate raw material[40]. Therefore, we attribute this fluorescence signal as coming from the Raman microscope objective lens, made of soda-lime silicate material, which is excited with the 785 nm line collected in the backscattering geometry. This parasitic signal becomes noticeable in the case of the gold film on flat silicon for the 785 nm excitation line because for this wavelength the gold film acts as a good reflector and backscatters laser light, as the reflectivity of gold is more than 95% at 785 nm. The fluorescence background signal has been subtracted from the Raman spectra to reveal clearly the superimposed graphene bands (Fig. S1).

**Results and Discussion**

*Graphene on nanostructured silicon*

Figure 2 shows scanning electron micrographs (SEM) of graphene supported on the uncoated nanostructured silicon surface (Figs. 2a and b) and on nanostructured silicon decorated with gold nanoparticles (Figs. 2c and d). The dashed lines in Figs. 2a and c indicate the edges of graphene. The nanostructured silicon surface, known as black silicon, is covered by a quasi-ordered array of columnar nanopillars (Figs. 2a) with a mean pillar diameter of 180 nm and a mean height of 860 nm, which are monolithically formed on silicon upon femtosecond laser irradiation in water, due to ultrafast melting and interference effects[41]. The mean distance between neighboring nanopillars is 320 nm center-to-center. Upon thermal evaporation of gold, the silicon nanopillars are decorated with aggregates of gold nanoparticles (Fig. 2d). The original CVD-grown graphene layer on copper is ~1×1 cm$^2$ but it tears to smaller pieces during the wet transfer process used in this work[39,42]. Still, there are large and uniform areas of graphene on the nanostructured silicon substrate, as shown in Fig. S2a (Supporting Information). Transferring graphene on rough substrates, such as nanostructured silicon, is more challenging compared with flat substrates because liquids can be trapped within the substrate nanostructure, resulting in graphene tears and cracks[43]. Figures 2a and c show transferred graphene layers, around 8×4 μm$^2$ and 13×7 μm$^2$ in area, respectively, supported by the columnar nanopillars of silicon and suspended in the gaps between them (Figs. 2b and d). Due to its flexibility, graphene conforms to the nanotopography of the substrates, increasing its interface area with the silicon nanopillars and gold nanoparticles. Additional SEM images can be found in the Supporting Information (Figs. S2 and S3).

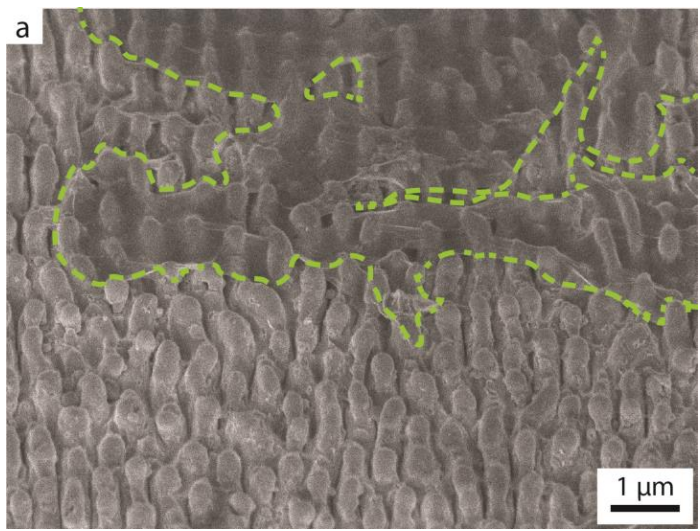

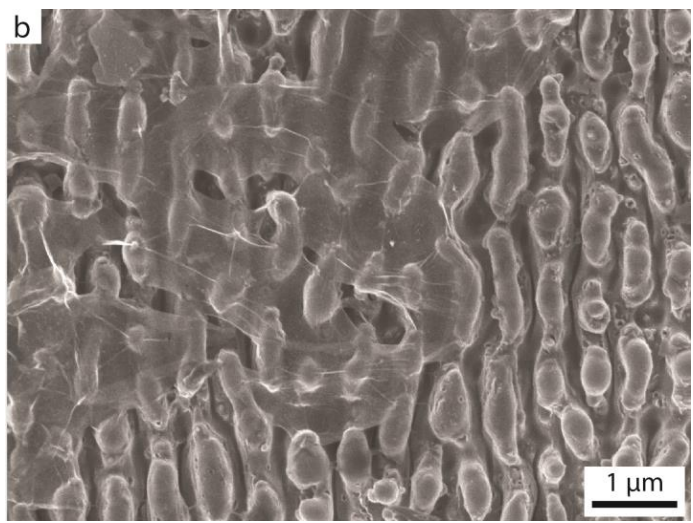

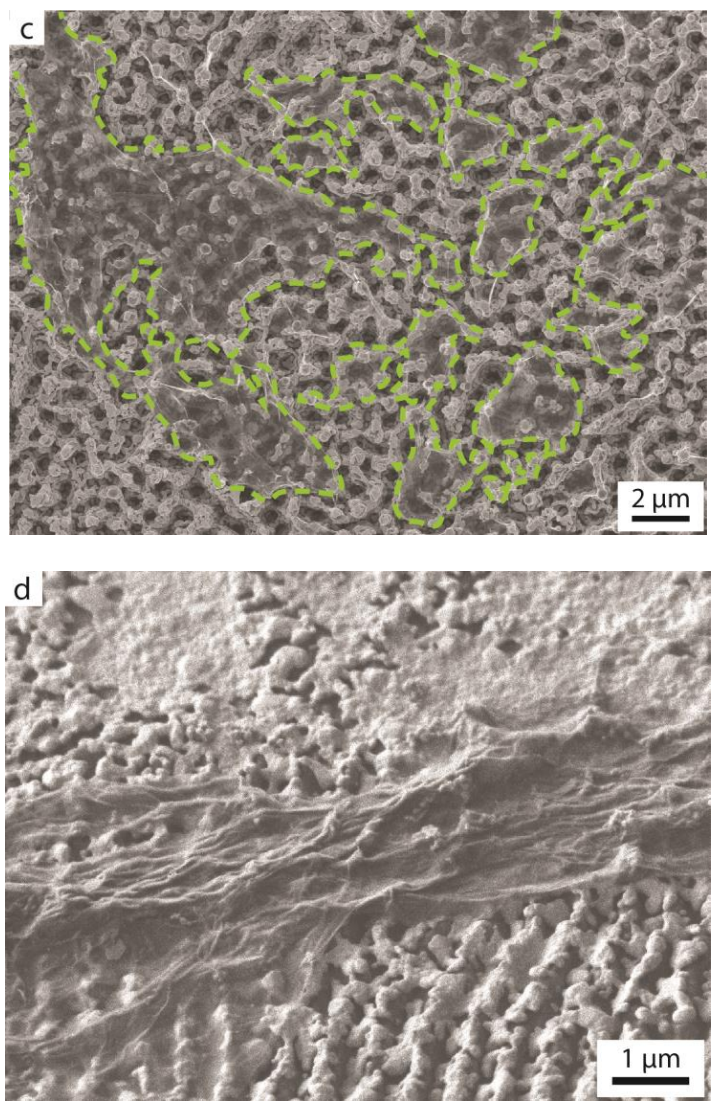

**Figure 2:** Scanning electron micrographs (SEM) of graphene on uncoated nanostructured silicon at (a) side (45°) view and (b) top view and on nanostructured silicon with gold nanoparticles at (c) top view and (d) side (45°) view. Dashed lines indicate the graphene layer.

The 3D plasmonic substrates, employed in this work for Raman spectroscopy of graphene, are developed by laser processing of silicon in water. This tabletop nanopatterning method is single-step, cost-effective, rapid, does not require the use of lithographic masks or vacuum, and is scalable as it poses no inherent limitation on the processed silicon area[44]. The nanostructured silicon substrates show great mechanical robustness as the nanopillars are monolithically formed on the original silicon wafer. The morphology of the surface is reproducible and controllable by tuning the fabrication parameters, such as laser pulse duration, surrounding medium, laser fluence, wavelength, and number of pulses[45]. Spontaneous formation of gold nanoparticles upon thermal evaporation of gold on nanostructured silicon further simplifies the fabrication process. The size and density of the nanoparticles depends on the amount (nominal thickness) of the metal coating, the deposition rate, and the substrate temperature[30]. Post-deposition processing, such as laser heating and melting of the metallic layer, leads to additional modifications of the nanoparticle morphology[34]. Thus, the optical properties of the plasmonic substrates may be tuned according to the application[29,30,34].

*Raman spectra of graphene on silicon substrates*

Figure 3 presents Raman spectra of graphene on four different substrates (flat pristine silicon, flat silicon with 50-nm gold film, uncoated nanostructured silicon, and nanostructured silicon with gold nanoparticles) with four different excitation wavelengths (488, 514, 633, and 785 nm). Because the Raman spectra of graphene on the plasmonic silicon substrate are dramatically enhanced, the spectra on the other substrates are shown magnified as depicted in Fig. 3 so that the main spectral features are visible. As the Methods section reports, all laser line spectra on the nanostructured substrates were acquired on the same sample location, for purposes of comparison. Especially for the plasmonic silicon substrate, the sample area shown in Fig. 2d is exactly the location where Raman spectra shown in Fig. 3 were acquired, for all excitation wavelengths. Apart from graphene peaks, the overtone (ca. 963 cm$^{-1}$) of the silicon optical phonon peak (520.5 cm$^{-1}$) is present for the uncoated silicon substrates and disappears for the gold-coated substrates for all excitation wavelengths. The G and 2D peaks of graphene are clearly detected on the Raman spectra of all substrates and they are greatly enhanced for the plasmonic substrate. The enhancement of the Raman scattering intensity on the plasmonic substrate allows the observation of the fine structure of the Raman spectrum of graphene. Indeed, we observe peaks which are typically difficult to detect in normal Raman scattering of graphene on pristine silicon substrates (see, for example, graphene on flat Si spectra in Fig. 3). Line scans across graphene pieces transferred on the substrates indicate that even though graphene tears into pieces, there is enough room for Raman scattering measurements over areas μm-size long (Supporting Information Figs. S4–S6). Additionally, line scans demonstrate the uniformity of the substrates, for which we obtain intensity ratios I(2D)/I(G) with a relative standard deviation (RSD) of 5.7% (Supporting Information Fig. S5).

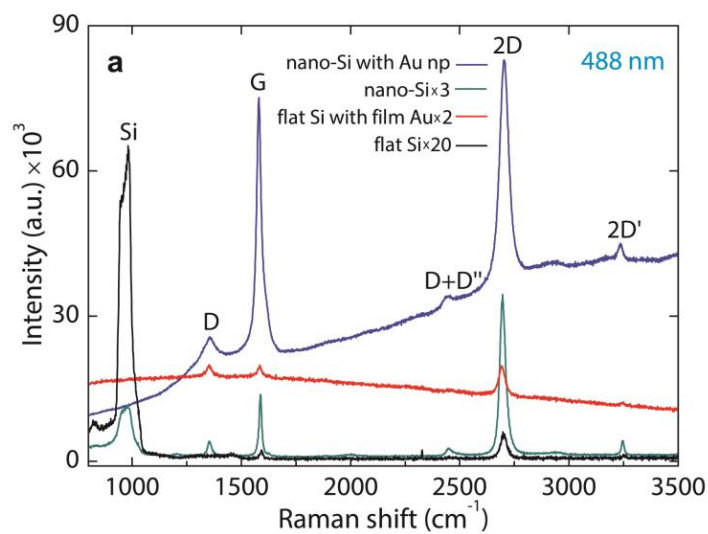

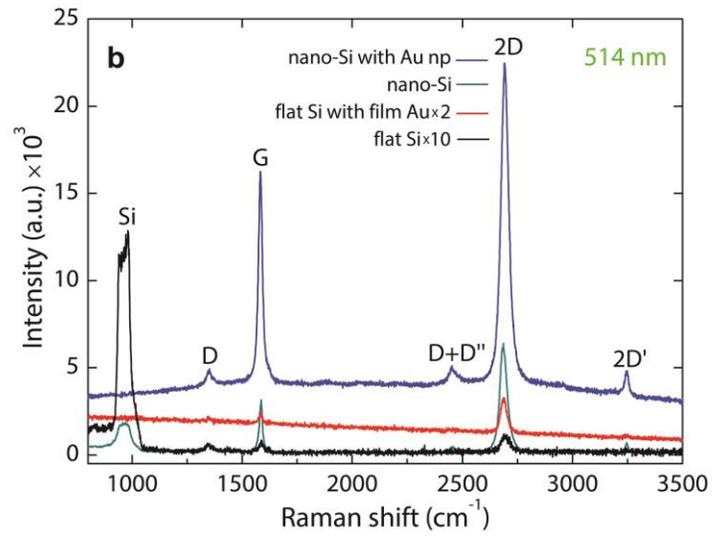
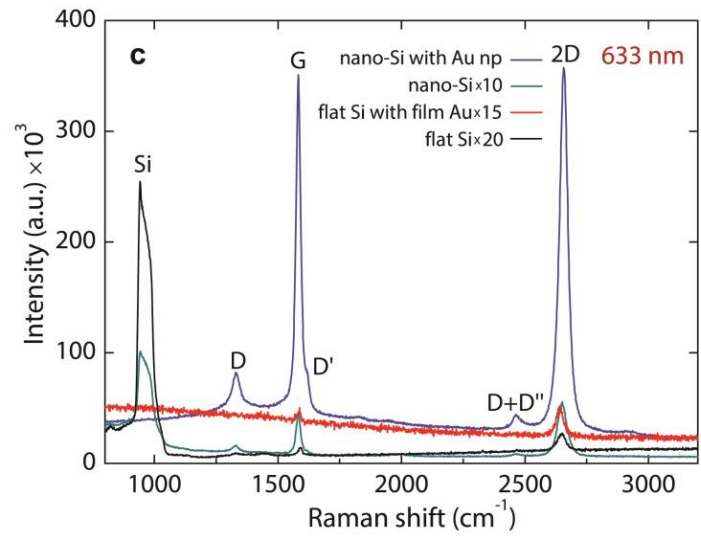

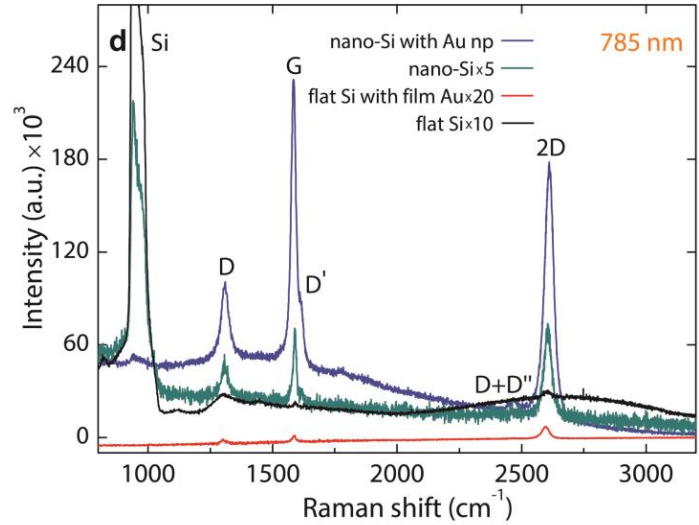

**Figure 3:** Raman spectra of graphene on flat pristine silicon, flat silicon with 50-nm gold film, uncoated nanostructured silicon, and nanostructured silicon with gold nanoparticles measured with (a) 488 nm, (b) 514 nm, (c) 633 nm, and (d) 785 nm excitation wavelengths. For clarity, spectra have been magnified by the indicated magnification factors, when necessary.

For all wavelengths, the D peak appears more intense for the nanostructured silicon substrates (coated and uncoated). This peak indicates the presence of defect states, which are necessary for its activation. Defects are introduced either during graphene growth or during the wet transfer process on the substrates. The low intensity or even absence, for some wavelengths, of the D peak on the flat silicon substrate indicates good quality growth. The enhancement of the D peak on the nanostructured substrates is probably due to folding and cracks induced in graphene during transfer on these nano-rough surface[22]. The D′ peak, also activated by the presence of defects in an intravalley process[14], appears near the G peak and it is often unresolved. In Fig. 3, the D′ peak is resolved only for the highly enhanced spectrum on the plasmonic substrate and appears clearly only for the longer excitation wavelengths of 633 nm and 785 nm. The D+D″ peak, assigned to a combination of a D phonon and a D″ phonon, also defect related, is detected only on the nanostructured silicon substrates (coated and uncoated), and, like the other peaks, it appears enhanced on the coated plasmonic nanostructured substrate. The D″ peak, expected at ~1100 cm$^{-1}$[14], is not observed. The overtones of D and D′ peaks, 2D and 2D′, respectively, do not require defects for their activation and are always present. In Fig. 3, the high-frequency 2D′ peak is shown only for the 488 nm and 514 nm excitation wavelengths, because for the 633 nm and 785 nm excitation wavelengths the Raman shift measured by the apparatus was limited to 3200 cm$^{-1}$, *i.e.*, below the 2D′ position. Similar to the other defect peaks, 2D′ is barely present for the flat silicon substrates and appears enhanced for the nanostructured silicon substrates.

The position of the G, D, and 2D peaks is shown in Fig. 4 as a function of the excitation photon energy, for graphene on each of the substrates used in this work. Lines are least square fits to the data. The D and 2D peaks are known to be dispersive with excitation energy[14,46]. Indeed, as shown in Fig. 4, these peaks are dispersive for all substrates employed here. For each peak, we fit separately the data for each substrate and then calculate the mean dispersion for all four substrates. For the D peak, we obtain a mean dispersion of 45.3± 2.4 cm$^{-1}$eV$^{-1}$ and for the 2D peak a mean dispersion of 97.7± 2.9 cm$^{-1}$eV$^{-1}$. These values agree with literature values[46-48]. Furthermore, Fig. 4 shows that the dispersion of the D and 2D peaks is similar for graphene on flat and nanostructured substrates, indicating that the morphology of the substrate does not affect this property of graphene. The G peak is known to be non-dispersive with excitation photon energy, as shown for all substrates in the inset of Fig. 4 [14].

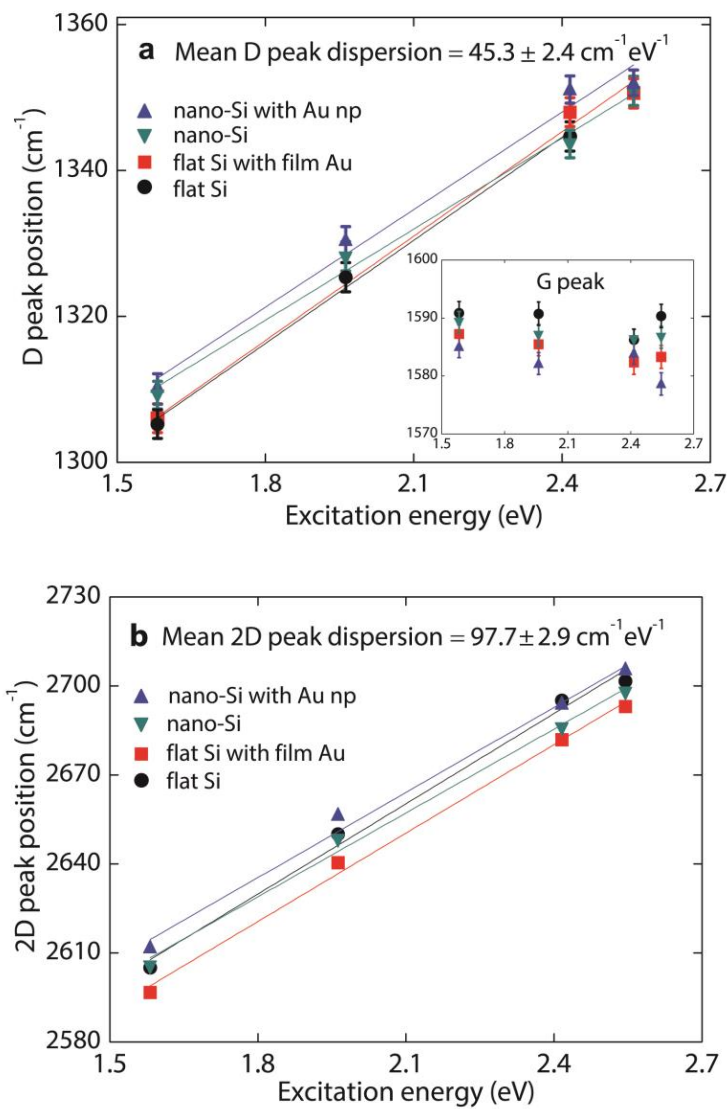

**Figure 4:** Position of (a) D peak and (b) 2D peak of graphene as a function of excitation photon energy for all substrates employed in this work. Inset in (a): G peak position. In Fig. 4b error bars are not visible due to the vertical axis scale.

The position of G and 2D peaks varies with the substrate. In Fig. 4 we observe blue-shifted G and 2D peaks for the flat pristine silicon substrate, compared with usual literature values of unstrained and undoped graphene[49-51]. This indicates the presence of p-doping in the graphene layer[13,52], induced either by the p-type silicon substrate or during graphene growth and transfer. For the uncoated nanostructured silicon substrate, both G and 2D peaks are red-shifted with respect to the flat silicon substrate (the effect appears more pronounced for the G peak position, which is plotted on a finer scale). This may be explained by the synergistic action of two effects: first, tensile strain, induced upon transfer of graphene on this non-planar substrate[13], and second, a reduced contact area of graphene with p-type silicon, as graphene is partially suspended between nanopillars on this substrate, which leads to a reduction of substrate-induced p-doping. The G and 2D peaks are even more red-shifted for the flat silicon substrate with gold film, with respect to the flat pristine silicon substrate. In this case, the peak positions are similar to the ones of unstrained and undoped graphene, indicating this substrate does not significantly modify the electric and mechanical properties of graphene. Finally, for the nanostructured silicon substrate with gold nanoparticles, the G peak is red-shifted and the 2D peak is blue-shifted compared with the flat silicon substrate. The behavior of graphene deposited on this substrate needs complex analysis as the gold nanoparticles induce doping[17,49], including the injection of hot carriers into which the surface plasmon resonance decays via Landau damping, and the morphology of the substrate induces tensile strain in graphene. Doping and strain configurations on this substrate are totally unknown and their interaction causes the simultaneous red shift of the G peak and the blue shift of the 2D peak. Strain and doping effects are also evident on the full width at half maximum (FWHM) of the Raman peaks, presented in Supporting Information (Fig. S7). A detailed investigation of the peak position and FWHM with respect to the substrate will be the subject of a future study.

In Figure 3 we observe that the Raman signal of graphene is enhanced on all substrates compared with the flat pristine silicon substrate, for most excitation wavelengths. To quantify the enhancement of Raman scattering we define the enhancement factor, $F = I_{\text{substrate}}/ I_{\text{flat Si}}$, as the ratio of the Raman signal intensity of selected graphene peaks on each substrate divided by the Raman signal intensity of graphene on the flat silicon substrate. Table 1 presents the measured enhancement factors for the G and 2D peaks of graphene for each substrate used in this work. A comparison of the experimental Raman enhancement factors with the results of simulations will be made in the following Section as a function of the excitation wavelength (see Fig. 9). As shown in Table 1, the highest enhancement factors, between two and three orders of magnitude, are obtained with the nanostructured silicon substrate with gold nanoparticles. For this substrate, the highest measured enhancement factor is $F = 880$ for the G peak and $F = 420$ for the 2D peak, both under 633 nm excitation wavelength. These values are similar to the highest measured enhancement factors reported in literature (~1000) for SERS of graphene on plasmonic substrates[17-28]. A higher enhancement is achieved when graphene is placed on top of plasmonic structures, as in this work, than when it is placed below them. This is because shadowing effects are avoided in the former case[25,28]. It should be noted that although the maximum apparent enhancement factor measured in this work is 880, the local enhancement factor may be much higher due to the fact that the Raman excitation beam diameter on the sample is 2–4 μm, averaging contributions from nanometric regions of the sample with strong Raman enhancement (hot spots) and regions with negligible contribution, as it is revealed below by the theoretical calculations shown in Fig. 8 [22,26]. The enhancement factors for the uncoated nanostructured silicon substrate are approximately one order of magnitude lower than the enhancement factors for the Au-decorated nanostructured silicon substrate and the enhancement factors for the flat silicon substrate with 50-nm gold film are 4–10 times lower than those for the uncoated nanostructured silicon. The flat silicon substrate with the gold film does not show an enhancement for the 785-nm excitation wavelength. Instead, for this wavelength, the intensity of the G and 2D peaks is lower than that on the flat silicon substrate (for an explanation, see numerical simulations that follow in Section 3.3).

| Excitation wavelength (nm) | Flat Si with film Au | | Nano-Si | | Nano-Si with Au nanoparticles | |
|---|---|---|---|---|---|---|
| | $F_G$ | $F_{2D}$ | $F_G$ | $F_{2D}$ | $F_G$ | $F_{2D}$ |
| **488** | 13 ± 1 | 11 ± 1 | 46 ± 3 | 40 ± 2 | 568 ± 30 | 162 ± 8 |
| **514** | 4.6 ± 0.7 | 10 ± 1 | 47 ± 6 | 61 ± 6 | 189 ± 18 | 182 ± 18 |
| **633** | 1.5 ± 0.5 | 2.4 ± 0.3 | 10.7 ± 0.6 | 6.3 ± 0.3 | 880 ± 35 | 422 ± 17 |
| **785** | 0.6 ± 0.1 | 0.8 ± 0.1 | 28 ± 5 | 26 ± 3 | 526 ± 60 | 355 ± 40 |

**Table 1:** Measured Raman intensity enhancement factors, $F$, for G and 2D peaks of graphene. Enhancement factors are calculated with respect to graphene on flat silicon.

*Simulations of the Raman enhancement factor for graphene on silicon substrates*

We complement the Raman measurements with numerical simulations in order to confirm the plasmonic nature of the Raman enhancement and gain insight on the effect of the substrate topography. We use the finite-difference time-domain (FDTD) method[53,54] following the procedure established in Ref. 24. We assume that the absorption at a particular point in graphene is proportional to the locally enhanced tangential field intensity and that the Raman emission from that point is proportional to the corresponding Stokes-shifted enhanced intensity. Specifically, for an incident field $|\mathbf{E}_{\parallel}^0(\mathbf{r},\omega)|$ at point **r** on suspended graphene and modulated fields $|\mathbf{E}_{\parallel}(\mathbf{r},\omega)|$ and $|\mathbf{E}_{\parallel}(\mathbf{r},\omega_s)|$ when graphene is on a substrate, we calculate the Raman signal enhancement as $S(\mathbf{r},\omega) = |\mathbf{E}_{\parallel}(\mathbf{r},\omega)|^2 |\mathbf{E}_{\parallel}(\mathbf{r},\omega_s)|^2 / |\mathbf{E}_{\parallel}^0(\mathbf{r},\omega)|^4$, where $\omega$ and $\omega_s$ are the incident

and the Stokes-shifted frequencies. In each calculation, only the electric field component parallel to the graphene basal plane is taken into account, as signified by the (||) subscript. For a flat substrate, $S$ is by symmetry the same for every point **r** and thus yields the final result for enhancement, while for graphene on a corrugated substrate an integration has to be performed over the entire illuminated area. In our calculations we only assume periodic cells and thus the integration is performed over the unit cell area.

We start with flat substrates, semi-infinite silicon and Au(50nm)/Si, as shown in Fig. 5. Taking the first substrate as the reference case, we plot the calculated G and 2D Raman enhancement factors, $F = S_{\text{Au/Si}}/S_{\text{Si}}$, along with the corresponding experimental points for bands G and 2D for the four wavelengths of interest. An overall good agreement is found in regard to the spectral dependence, although there is some scattering in the data for the smaller wavelengths. The reason for the large enhancement at small wavelengths is the very strong quenching in the bare silicon case: simply, for light reflected off a flat surface of a semi-infinite material of complex refractive index $n$, the reflected amplitude is $r = (1 - n)/(1 + n)$ and the corresponding total complex field at the surface (where graphene would be) is $E = 1 + r = 2/(1 + n)$. Assuming graphene on the surface, its absorption is proportional to $\sigma|E|^2$ (where $\sigma$ is the in-plane graphene conductivity), the relative enhancement of absorption compared to graphene suspended in air is $4/|1 + n|^2$ and the relative Raman enhancement is $16/|1 + n|^4$, where we assumed for simplicity zero Stokes shift (we also note that the above considerations are approximate since the presence of graphene on the surface will slightly modify the field values). The graphene absorption enhancement, expressed as the ratio of absorption of graphene on gold divided by absorption of graphene on silicon, is also plotted in Fig. 5. Using literature data for the refractive index of silicon[55] and gold[56] substrates (assuming for simplicity the latter to be semi-infinite instead of 50 nm), we get at 500 nm incident wavelength ($n_{Si} \cong 4.3 + i0.07$, $n_{Au} \cong 0.97 + i1.87$) $S_{Si} \cong 0.02$ and $S_{Au} \cong 0.28$, yielding $S_{Au}/S_{Si} \cong 14$, explaining the simulation results and measurements. At longer wavelengths, Au becomes more reflective and the actual field magnitude on the surface diminishes. For example, at 650 nm wavelength ($n_{Si} \cong 3.85 + i0.0165$, $n_{Au} \cong 0.156 + i3.6$) the Raman enhancement is estimated at $S_{Au}/S_{Si} \cong 2.7$, while at 800 nm wavelength ($n_{Si} \cong 3.69 + i0.0065$, $n_{Au} \cong 0.154 + i4.9$) we get a reduction $S_{Au}/S_{Si} \cong 0.75$. However, we must note that in all cases there is a quenching compared to the Raman signal expected from graphene suspended in air. In light of this, we anticipate a significant increase in signal once we remove graphene away from the flat substrate, *e.g.*, as in placing it on top of nanostructured silicon.

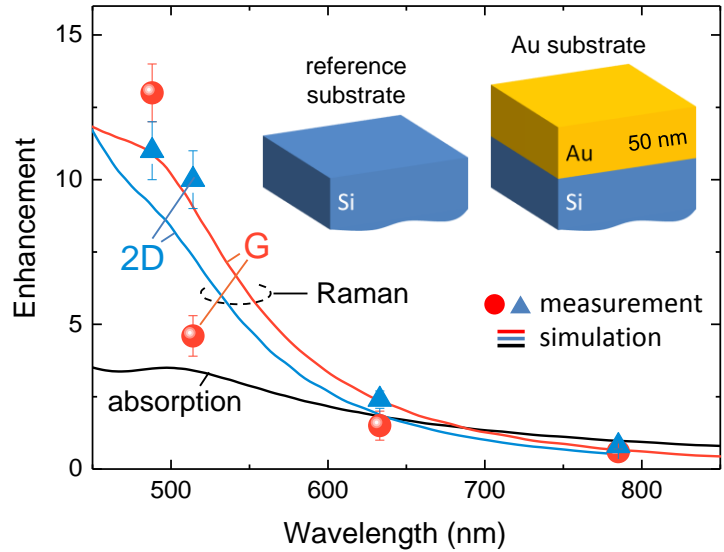

**Figure 5:** Measured and simulated Raman enhancement factors for graphene on Au(50nm)/Si compared to graphene on silicon. The calculated graphene absorption enhancement is also plotted for reference. Solid red and blue lines denote simulation results and symbols denote the corresponding experimental data.

For nanostructured silicon we assume a simplified model consisting of a periodic square array of silicon pillars with period $L = 400$ nm, base radius $\rho_0 = 150$ nm and height $z_t = 500$ nm (Fig. 6a). The pillars follow the functional form $z = z_t(1 - \rho^4/\rho_0^4)$, where $\rho = \sqrt{x^2 + y^2}$ is the radial position away from the pillar center. These parameters are close to the silicon nanopillar structure observed in the SEM images in Fig. 2. For the graphene conformation on top of the nanostructured silicon we assume a two-dimensional functional $z = z_t f(x) f(y)$ with $f(x) = 1 - a\left|\cos^d\left(\frac{\pi x}{L}\right)\right|$ (similarly for y), where $a = 1 - \sqrt{1 - w/z_t}$ and $w$ is the maximum depth of the graphene sheet drop in-between the pillars ($w \leq z_t$, see Fig. 6a). The exponent $d$ determines the shape of the graphene sheet at the top of the pillar and is adjusted in order to better match the functional form of the pillar itself at its top. In our calculations we used $d = 4$ for the uncoated nanostructured silicon case and $d = 6$ for the Au-decorated nanostructured silicon. Different graphene conformations for various $w$ values are shown in Fig. 6a by dashed lines. Our goal is to calculate the Raman enhancement as a function of $w$ and gain insight into the effect of graphene conformation. To avoid performing an enormous amount of calculations, we take the simplified approach of performing the simulation without graphene and monitoring the modulated fields at every point in the structure. Then, in post-processing mode we integrate the tangential product $|\mathbf{E}_{||}(\mathbf{r}, \omega)|^2 |\mathbf{E}_{||}(\mathbf{r}, \omega_s)|^2$ over the presumed graphene sheet area (note that the graphene area now is larger than $L^2$ due to conforming around the silicon pillars). The parallel component is found by $\mathbf{E}_{||}(\mathbf{r}) = \mathbf{E}(\mathbf{r}) \times \hat{\mathbf{n}}(\mathbf{r})$, where $\hat{\mathbf{n}}(\mathbf{r})$ is the graphene surface normal unit vector at position $\mathbf{r}$ ($\hat{\mathbf{n}} \equiv \left(\frac{\partial z}{\partial x}, \frac{\partial z}{\partial y}, -1\right)$, $z = z_t f(x) f(y)$ is the two-dimensional functional defining the graphene conformation, see above).

The electric field intensity distribution, $|\mathbf{E}(\mathbf{r},\omega)|^2/|\mathbf{E}_0(\mathbf{r},\omega)|^2$, is shown on a vertical crosssection for the four incident wavelengths of interest in Fig. 6b. The Fabry-Perot standing waves, which of course vary for the different wavelengths, give an indication of the importance of the graphene conformation, particularly on the value of the *w* parameter in our model. The Raman enhancement factors with respect to graphene on flat silicon are shown in Figs. 6c,d for the G and 2D bands, respectively, for the four wavelengths of interest. Strong fluctuations are observed as the graphene depth *w* increases. The variation in 2D is smaller than the one for G, which is however expected taking into account that $\omega_s$ is more red-shifted in the former case. With dashed horizontal lines denoting the experimental measurements, we note a good qualitative agreement for three excitation wavelengths (488, 514, and 785 nm) but a rather large deviation for 633 nm. Theoretically, we do not expect such a small enhancement value at this particular wavelength and, thus, cannot explain this disagreement at this point. Aside from that and considering the large number of simplifications made in this theoretical study regarding the actual silicon nanostructure (random pillar heights, thicknesses, shapes, and separations) and the actual graphene conformation on it, the agreement concerning the order of magnitude of the enhancement is more than satisfactory.

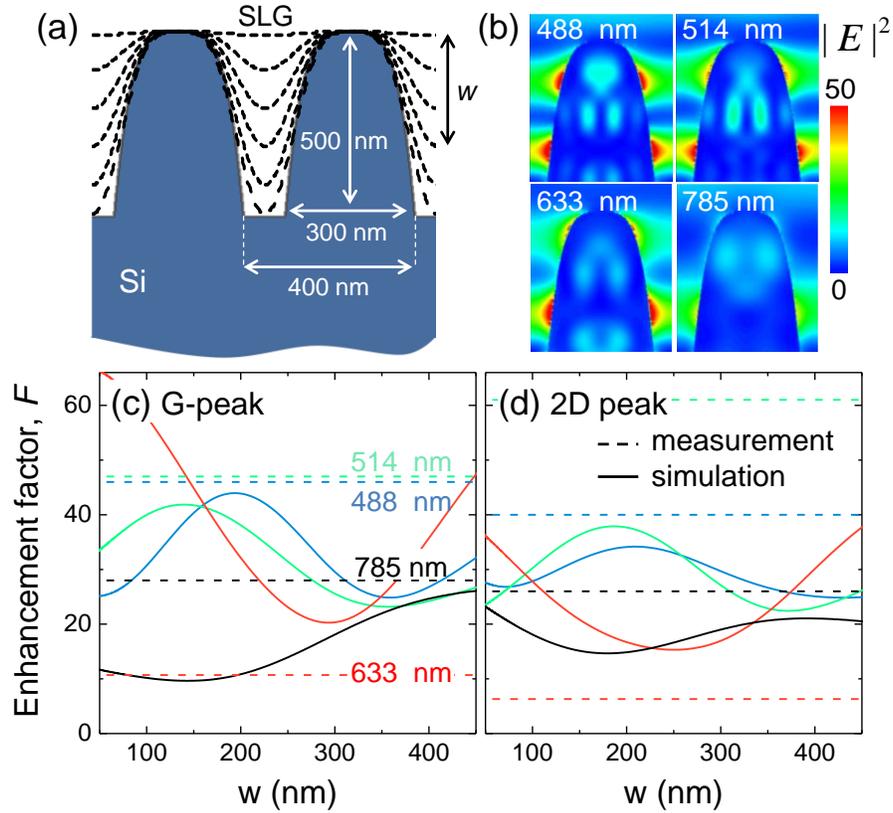

**Figure 6:** (a) Simulation model consists of a square array of silicon pillars. (b) Electric field intensity enhancement distribution (compared to the incident intensity) for the four wavelengths of interest. (c) G and (d) 2D Raman peak enhancement factor (with respect to graphene on flat silicon) as a function of the maximum graphene conformation depth, $w$, in-between the silicon pillars. Dashed lines indicate the experimental measurements.

Next, we simulate the Au-decorated nanostructured silicon substrate. A cross section of the model is shown in Fig. 7a, where 50-nm diameter gold spheres have been placed on the silicon pillars. Here, we ignore the presence of a native oxide silicon layer on the silicon pillars because its effect on the electromagnetic response of the substrate is negligible (Figs. S8 and S9 in Supporting Information). The graphene sheet conformation is now also around the gold spheres. To keep our model simple, we assume again the same functional form for graphene, but with its highest point to be through the nanoparticle center, *i.e.*, $z_t = 525$ nm. This effectively assumes that graphene wraps around half of the nanoparticle. The Raman calculation follows the same process as before, utilizing the calculated field intensity distribution, whose profile for the four wavelengths is shown in Fig. 7b. Due to the Fabry-Perot resonances, now different nanoparticles become more active for different wavelengths. Figs. 7c,d show the estimated Raman enhancement factors as a function of *w* with respect to graphene on flat silicon. The experimental results are denoted by the horizontal dashed lines. Rather than an oscillatory behavior due to standing wave resonances, as in the case of uncoated nanostructured silicon, we note in the simulation results a monotonic increase of the Raman enhancement with increasing graphene conformation depth, *w*. This is due to the graphene sheet sampling more plasmonic nanoparticle near fields and thus emitting a stronger SERS signal. Additionally, as shown in Fig. 7b, nanoparticles lying near the base of the silicon pillars create stronger electric near fields, compared with nanoparticles lying near the top of the pillars. In our calculations, we ignore the quenching of the plasmonic response of the gold nanoparticles due to graphene absorption, as this is negligible for single-layer graphene (Fig. S10 in Supporting Information). There is again some scatter between the experimental and simulation results, but this should be expected considering the added complexity of the gold nanoparticle decorations (size, shape, and distribution) combined with the complexity of the nanostructured silicon substrate. Variations in the gold nanoparticle size, present in this study due to the spontaneous formation of the nanoparticles, are expected to cause small changes in the distance of graphene from the silicon pillars. However, the plasmonic response of the nanoparticles heavily dominates the Raman enhancement, as shown by comparing Figs. 6 and 7, therefore the effect of the distance of graphene from the silicon pillars is not noticeable. The actual Raman hot spots on the graphene sheet, projected on a horizontal plane, are depicted in Fig. 8 for $\lambda =633$ nm and $w =450$ nm. The strong influence of the plasmonic near fields is evident. At the Raman hot spots the local enhancement is calculated to reach values as high as 2500 and 1600 for the G and 2D peaks, respectively. Simulations of graphene on flat silicon decorated with gold nanoparticles (Supporting Information Figs. S11 and S12) show that the SERS enhancement of graphene is an order of magnitude less than the enhancement on Au-decorated silicon pillars, which is due to destructive interference between incident waves and waves reflected off the flat silicon interface. Therefore, the 3D topography, which disrupts the flat interface, is preferential for SERS on reflective surfaces.

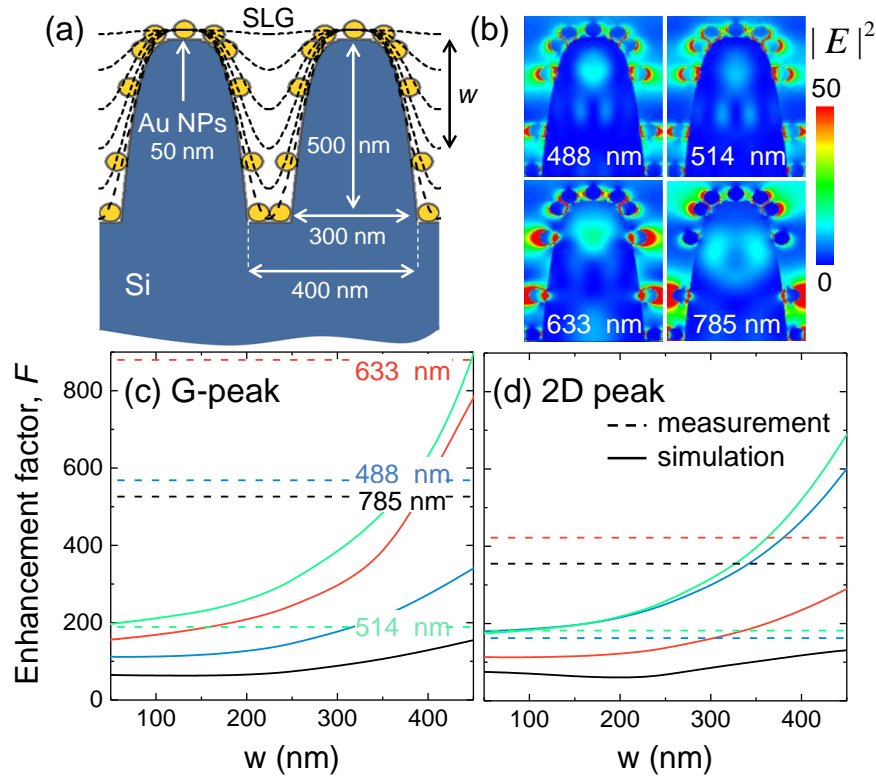

**Figure 7:** (a) Simulation model consists of a square array of silicon pillars decorated with gold nanoparticles. (b) Electric field intensity enhancement distribution (compared to the incident intensity) for the four wavelengths of interest. (c) G and (d) 2D Raman peak enhancement factor

(with respect to graphene on flat silicon) as a function of the maximum graphene conformation depth, *w*, in-between the silicon pillars. Dashed lines indicate the experimental measurements.

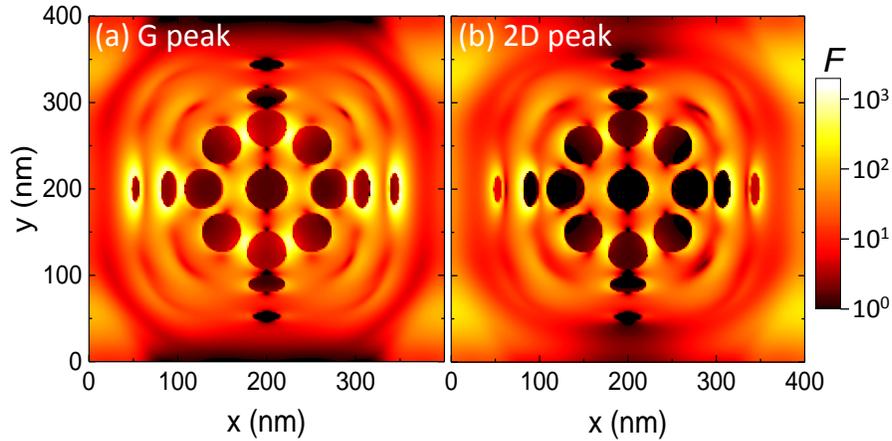

**Figure 8:** The distribution of Raman (a) G and (b) 2D peak enhancement factor of graphene on nanostructured silicon with gold nanoparticles (with respect to the flat silicon case) projected on to a horizontal plane for incident wavelength 633 nm and graphene conformation depth $w = 450$ nm.

To summarize our comparison between measurements and simulations, we average the calculated results for all *w* values and plot them along with the experimental values in Fig. 9. The vertical axis (Raman enhancement factor) is in logarithmic scale so all three substrate cases are shown in the same graph. In terms of order of magnitude, a good overall agreement is found in all cases. We can confidently conclude that there is a clear plasmonic effect separating the decorated nanostructured silicon case from all the others. The scale of separation is about one order of magnitude. Also, we note that the reference case here is graphene on flat silicon. From our estimates in the beginning of the theoretical considerations, we expect a quenching of the order of 40 compared to graphene suspended in air (actual numbers for quenching are 50, 35, and 30 for 500, 650, and 800 nm wavelength, respectively). This puts the uncoated nanostructured silicon results on par with suspended graphene: it is far from the highly-quenching flat silicon surface and the Fabry-Perot resonances only slightly modulate the actual final signal enhancement. The higher quenching for flat silicon at small wavelengths (Fig. 5) results into the higher enhancement values compared to the ones at longer wavelengths, as seen in Fig. 9, middle part. For the decorated nanostructured silicon, on the other hand, the Raman signal is clearly above the one from suspended graphene and definitely enhanced due to plasmonic fields. As a final comment, we note that while the scattering of measured and simulated data is expected given the large difference between simulated (periodic) and actual (random) substrate structures, the good overall agreement in terms of trends and order of magnitude is promising. For example, making measurements on periodic substrates we could potentially use simulations to estimate the actual graphene conformation. Such periodic structures would also allow for controlling the period and width of the silicon nanopillars, which in turn would affect the enhancement of the graphene Raman signal, however they are beyond the scope of this work, which focuses on the simplicity and scalability of laser processing.

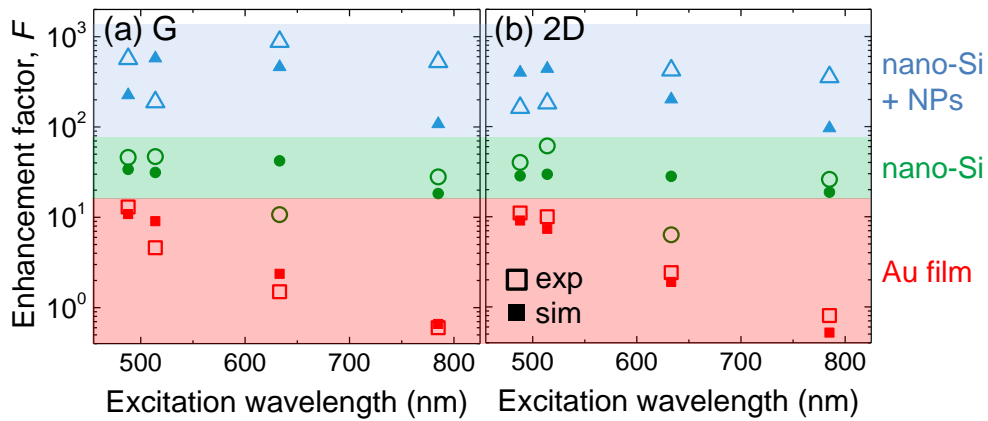

**Figure 9:** Summary comparison between the experimental (open symbols) and simulation (solid symbols) results for the (a) G and (b) 2D peak Raman enhancement factors for graphene on flat

Au (squares), uncoated nanostructured silicon (circles), and nanostructured silicon with gold nanoparticles (triangles), with respect to graphene on flat silicon.

Figure 9 shows that there is not a strong variation of the enhancement factor with excitation wavelength, even for the plasmonic nanostructured silicon substrate with gold nanoparticles. Indeed, this substrate provides significant enhancement over a broad range of excitation wavelengths, spanning the visible electromagnetic spectrum. The formation of aggregates of gold nanoparticles on the surface of this substrate leads to broadening of the plasmon resonance over the visible range. Additionally, broadening of the plasmon resonance is induced by the variation of the dielectric environment sampled by each nanoparticle on the silicon nanopillars, as shown in Fig. 7a. Nanoparticles on pillars are no longer indistinguishable, like nanoparticles on a flat substrate, as the relative orientation of the interface area of each nanoparticle with silicon and with graphene varies according to its location on the pillar (Figs 7b and 8). The broad range of excitation wavelengths for which significant enhancement is observed renders the coated nanostructured silicon substrates flexible for use with a wide range of Raman spectrometers, employing various laser sources for multiple applications[30].

**Conclusions**

We have integrated, for the first time, graphene with a 3D, non-planar, scalable platform, compatible with the vast silicon infrastructure and monolithic electronic/optoelectronic systems. The mechanical flexibility and strength of graphene, and 2D materials in general, allow them to conform to the topography of the substrate. We have probed the effect of the nanotopography of the silicon platform with gold nanoparticles (plasmonic substrate) on the interaction of graphene with electromagnetic radiation via the SERS case. The 3D geometry we propose has two advantages compared with 2D plasmonic geometries, achieving near-record enhancement of the Raman signal of graphene: a) due to its flexibility, graphene conforms to the 3D substrate nanotopography, increasing its interface area with plasmonic nanoparticles and the sampling of enhanced near fields and b) even for uncoated, bare silicon nanopillars, graphene adopts a semi-suspended topography between them, which results in enhanced electromagnetic interaction and Raman signal, similar to this of suspended graphene in air. Additionally, broadband enhancement of the light fields across the visible spectrum is achieved in contrast with most plasmonic substrates which show narrow resonances. Our theoretical analysis reveals the broadband enhancement is due to the unique morphology of the 3D substrates. This morphology leads to the formation of aggregates of metallic nanoparticles on the Si nanopillars, broadening the plasmon resonance, and also to synergistic effects between the nanoparticles and the dielectric environment of the nanopillars, which render different nanoparticles more active for different wavelengths and locations on the pillars. The maximum enhancement (×880 for the G peak), obtained for 633 nm excitation, is similar to the highest measured enhancements reported in the literature for SERS of graphene on plasmonic substrates. FDTD numerical simulations revealed local enhancement factors as high as 2500 and demonstrated a monotonic increase of enhancement with the degree of graphene conformation to the 3D topography of the substrate.

Combinations of strain and doping in graphene were evident on the Raman spectra obtained on the different substrates employed in this work. Even though the transfer process does not leave the original CVD-grown graphene layer intact, smaller layers of graphene are formed, which maintain the properties of single-layer graphene. Advanced transfer processes may also be combined with the 3D substrates employed here, to prevent cracks and tears of graphene[43].

Because these substrates are known to provide spatially uniform SERS enhancement factors, graphene tearing is not a severe limitation for Raman spectroscopy with them and requirements on the transfer process can be relaxed. Optoelectronic devices will benefit tremendously from hybrid nanostructured silicon-graphene building blocks with high responsivity due to the large surface area of the nanostructure, fast response speeds due to the reduced dimensions of the materials, and spectrally broad photoresponse. The broadband plasmonic enhancement of the proposed platforms is also beneficial for ultrasensitive SERS-based sensors operating with a variety of laser sources. The ease of fabrication, scalability, and excellent electromagnetic enhancement of laser-structured plasmonic silicon platforms, along with their successful integration with graphene, which may also be extended to most 2D materials and their van der Waals heterojunctions, pave the way for future real-world applications of large-area 2D devices with complex functionalities in the fields of sensing, photonics, optoelectronics, and medical diagnostics, among others.

**SUPPORTING INFORMATION**

We provide a file of supporting information including additional SEM images, Raman analysis, and simulations.

Supporting Information_SERSgraphene (PDF)

AUTHOR INFORMATION

**Corresponding Author**

* Email: kandyla@eie.gr

ACKNOWLEDGMENT

**Funding Sources**

We acknowledge support of this work by the EU Graphene Flagship (no. 696656) and by the project "Advanced Materials and Devices" (MIS 5002409) which is implemented under the "Action for the Strategic Development on the Research and Technological Sector", funded by the Operational Program "Competitiveness, Entrepreneurship and Innovation" (NSRF 2014-2020)and co-financed by Greece and the European Union (European Regional Development Fund). M. Kanidi acknowledges support through a Ph.D. fellowship by the General Secretariat for Research and Technology (GSRT) and the Hellenic Foundation for Research and Innovation (HFRI).

We thank Dr. Doris Möncke, scientific external collaborator of TPCI/NHRF, for pointing out to us reference [40].

ABBREVIATIONS

SERS, surface-enhanced Raman spectroscopy; CVD, chemical vapor deposition; SEM, scanning electron microscopy; FDTD, Finite-difference time-domain; CCD, charge coupled device; FWHM, full width at half maximum.